\documentclass[aps,prl,twocolumn,superscriptaddress,showpacs]{revtex4}

\usepackage{graphicx}

\begin{document}


\title{Supermagnetosonic Jets behind a Collisionless Quasi-parallel Shock}



\author{H.~Hietala}
\email[]{heli.hietala@helsinki.fi}
\affiliation{Department of Physics, University of Helsinki, Helsinki, Finland}

\author{T.~V.~Laitinen}
\affiliation{Swedish Institute of Space Physics, Uppsala, Sweden}

\author{K.~Andr\'eeov\'a}
\affiliation{Finnish Meteorological Institute, Helsinki, Finland}

\author{R.~Vainio}
\affiliation{Department of Physics, University of Helsinki, Helsinki, Finland}

\author{A.~Vaivads}
\affiliation{Swedish Institute of Space Physics, Uppsala, Sweden}

\author{M.~Palmroth}
\affiliation{Finnish Meteorological Institute, Helsinki, Finland}

\author{T.~I.~Pulkkinen}
\affiliation{Finnish Meteorological Institute, Helsinki, Finland}

\author{H.~E.~J.~Koskinen}
\affiliation{Department of Physics, University of Helsinki, Helsinki, Finland}
\affiliation{Finnish Meteorological Institute, Helsinki, Finland}

\author{E.~A.~Lucek}
\affiliation{The Blackett Laboratory, Imperial College, United Kingdom}

\author{H.~R\`eme}
\affiliation{Centre d'Etude Spatiale des Rayonnements (CESR), University of Toulouse, UPS, Toulouse, France}
\affiliation{CESR/CNRS, Toulouse, France}




\date{November 2, 2009}

\begin{abstract}

The downstream region of a collisionless quasi-parallel shock is structured containing bulk flows with high kinetic energy density from a previously unidentified source. We present Cluster multi-spacecraft measurements of this type of supermagnetosonic jet as well as of a weak secondary shock front within the sheath, that allow us to propose the following generation mechanism for the jets: The local curvature variations inherent to quasi-parallel shocks can create fast, deflected jets accompanied by density variations in the downstream region. If the speed of the jet is super(magneto)sonic in the reference frame of the obstacle, a second shock front forms in the sheath closer to the obstacle. Our results can be applied to collisionless quasi-parallel shocks in many 
plasma environments.

\end{abstract}



\maketitle




\textit{Introduction.---}
When the angle between the nominal shock normal and the upstream magnetic field is small, the shock transition in a collisionless plasma is much more complex than in the quasi-perpendicular case \cite{Burgess05}.
The nonthermal nature of the upstream side of a quasi-parallel shock has been recognized for decades \cite{Asbridge68,Hoppe82,Onsager90}.
 The downstream region, however, has only recently come under active research, both in astrophysical (supernovae \cite{Giacalone07}) and Solar System (termination shock \cite{Jokipii08Nature}, Earth's bow shock \cite{Retino07,Yordanova08}) contexts.

The most detailed and extensive data of collisionless shock waves are from the Earth's bow shock.
In contrast to remote observations and laboratory measurements, the near-Earth space can be used to study \emph{in situ} supersonic plasma flow past a magnetic obstacle---the flow of the solar wind around the magnetosphere of the Earth. The magnetospheric boundary (the magnetopause) is usually located at a distance of 10 Earth radii (1~$R_\mathrm{E}$~=~6371~km) in the solar direction. The bow shock is curved at magnetospheric scales while the structures in the solar wind and interplanetary magnetic field are large compared to the size of the magnetosphere. Hence the locations of parallel and perpendicular regions of the bow shock vary depending on the direction of the interplanetary magnetic field. Consequently, we can access a wide range of plasma conditions via spacecraft observations.

Recent measurements have shown that the flow in the downstream region of a quasi-parallel shock is structured:
Nemecek \textit{et al.} \cite{Nemecek98} have reported observations of transient ion flux enhancements in the Earth's magnetosheath during radial interplanetary magnetic field.
In subsequent studies, Savin \textit{et al.} \cite{Savin08} have found more than 140 events of anomalously high energy density.
However, the source of these jets of high kinetic energy and ion flux has remained unclear.
In this Letter, we present a set of multi-spacecraft measurements from Cluster \cite{Cluster} that allows us to suggest a formation mechanism for such jets.


\textit{Data.---}
We have analyzed Cluster measurements from the evening of March 17, 2007, when the four spacecraft (C1-C4) were close to the nose of the magnetosphere.
The spacecraft constellation was quite flat in the nominal plane of the magnetopause, since C3 and C4 were close to each other (950~km, 0.15~$R_\mathrm{E}$ apart), while the others were slightly more than 7000~km ($>$1~$R_\mathrm{E}$) away.
We have used data from the magnetic field experiment FGM from all four spacecraft, and from the ion experiment CIS-HIA from C1 and C3 \cite{Cluster}.
Information about the upstream conditions was provided by ACE and Wind satellites situated near the Lagrangian point L1, as well as the Geotail spacecraft, which at the time was in the foreshock region near the subsolar point.

\begin{figure}
\includegraphics[width = 0.45\textwidth]{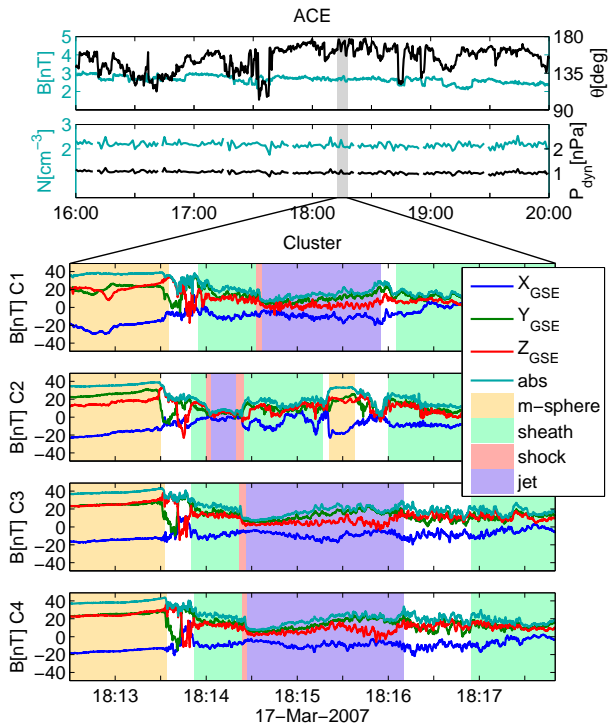}
\caption{Upper panels: upstream solar wind data from the ACE satellite (time-shifted by 44 minutes to account for the solar wind propagation to the magnetopause). First panel: magnitude of the interplanetary magnetic field and angle $\theta$ between the field direction vector and the $X_\mathrm{GSE}$-axis [22]. Second panel: plasma number density and dynamic pressure. ACE was located at $(X,Y,Z)_\mathrm{GSE}$ = (237, 36.4, $-$18.6)~$R_\mathrm{E}$. The gray shading marks the period of interest. Lower panels: magnetic field from all four Cluster spacecraft in GSE coordinates. The quartet was situated around (10.7, 1.5, 3)~$R_\mathrm{E}$. The color panels mark different plasma regions. White background between color panels represents transition between two regions.\label{fig:B}}
\end{figure}

The free upstream solar wind flow was quite fast ($V~\sim$~530~km/s) and steady (see the upper panels of Figure~\ref{fig:B}). The particle number density was around 2~$\mathrm{cm}^{-3}$ and hence the dynamic pressure ($\rho V^2$, where $\rho$ is the mass density) was low, close to 1~nPa. The interplanetary magnetic field was approximately radial, i.e., the sunward magnetic field component $B_X$ 
\footnote[21]{Geocentric solar ecliptic system (GSE) coordinates: $X$-axis points from the Earth towards the Sun, $Y$-axis opposite to the planetary motion, and $Z$-axis towards the ecliptic North Pole.}
was dominant. Moreover, the angle between the flow direction and the magnetic field was less than 20$^{\circ}$. Consequently, the bow shock was quasi-parallel at the subsolar point. The upstream Mach numbers
\footnote[22]{Characteristic speeds in plasma are Alfv\'en speed $V_\mathrm{A}~=~(B^{2}/\mu_0\rho_m)^{1/2}$, sound speed $V_\mathrm{S}~=~(\gamma k_{B} T/m)^{1/2}$, and magnetosonic speed $V_\mathrm{MS}~=~(V_\mathrm{A}^{2} + V_\mathrm{S}^{2})^{1/2}$. Thus the relevant Mach numbers are Alfv\'en Mach number $M_\mathrm{A}~=~V_{n}/V_\mathrm{A}$, sonic Mach number $M_\mathrm{S}~=~V_{n}/V_\mathrm{S}$, and magnetosonic Mach number $M_\mathrm{MS}~=~V_{n}/V_\mathrm{MS}$. Here $V_{n}$ is the component of velocity parallel to the shock normal in the reference frame moving with the shock front.}
 were all larger than 10: ${M_\mathrm{A}}$~$\sim$~12, ${M_\mathrm{S}}$~$\sim$~16, and ${M_\mathrm{MS}}$~$\sim$~10.
The location of the bow shock, as observed by Geotail at $(X,Y,Z)_\mathrm{GSE}$~=~$\sim$(14,~$-$7,~$-$3)~$R_\mathrm{E}$ [22] when the shock moved over the spacecraft several times between 17:30 and 24:00~UT, matches well to the empirical model~\cite{Merka05} for the measured upstream parameters.

%

The Cluster quartet, moving on an outbound orbit near the subsolar point, encountered the magnetopause the first time shortly after 17:00~UT and passed into the solar wind at 22:30~UT. Between 17:00 and 20:00~UT Cluster observed multiple magnetopause crossings.
Moreover, during this 3-hour period Cluster observed several high speed jets ($V \sim$ 500~km/s) in the magnetosheath behind the quasi-parallel bow shock. Here we concentrate on the jet between 18:13 and 18:17~UT.

\begin{figure}
\includegraphics[width = 0.45\textwidth]{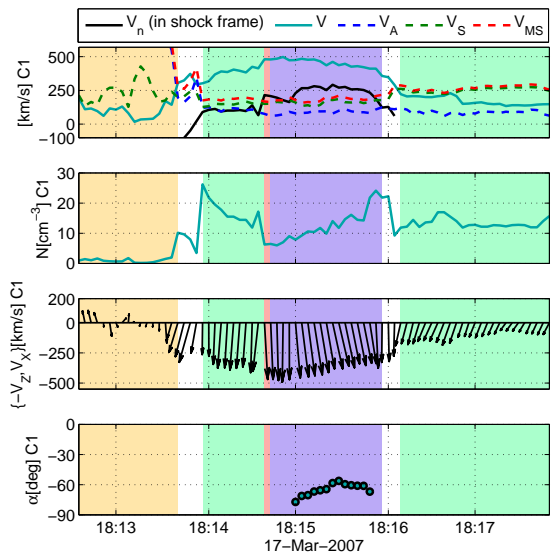}
\caption{First panel: Component of plasma velocity parallel to the local shock normal ($-$0.59, 0.52, $-$0.61)$_\mathrm{GSE}$ of the secondary shock (calculated with minimum variance analysis), in the reference frame moving with the secondary shock $V_{n}$ (black solid curve) and total plasma speed $V$ (light blue solid curve). Dashed curves show the characteristic speeds: Alfv\'en speed $V_\mathrm{A}$ (blue), sound speed $V_\mathrm{S}$ (green), and magnetosonic speed $V_\mathrm{MS}$ (red) [23].
Second panel: Plasma number density.
Third panel: Bulk velocity projection to ($-Z_\mathrm{GSE}$, $X_\mathrm{GSE}$) plane.
Fourth panel: The angle $\alpha$ calculated from the observed velocity deflection using the jump conditions for high ${M_\mathrm{A}}$ and $r=4$. The calculation is not expected to be valid at the edges of the jet where the shock is weak, and hence $\alpha$ is shown for the center only.
All data are from C1. 
The color coding for different plasma regions is the same as in Figure~\ref{fig:B}.
\label{fig:V}}
\end{figure}

As displayed in the lower panels of Figure~\ref{fig:B}, all four spacecraft were inside the magnetosphere at the beginning of the interval.
First the magnetopause moved inwards passing over the Cluster quartet at 250~km/s (obtained using 4-spacecraft timing).
Then the spacecraft observed a weak shock within the magnetosheath moving in the same direction at 140 km/s. In the first panel of Figure~\ref{fig:V}, the C1 measurements show that at this moment the component of the plasma velocity parallel to the shock normal in the reference frame moving with the shock $V_{n}$ exceeds the magnetosonic speed $V_\mathrm{MS}$ (as well as the other characteristic speeds \footnotemark[23]). Hence the magnetosonic Mach number $M_\mathrm{MS} > 1$. (This is also the case with respect to the magnetopause.) The same was observed by C3 located about 8000~km away (not shown).

After the shock Cluster entered a cold, supermagnetosonic jet with a plasma speed close to 500 km/s (see Figure~\ref{fig:V}, first panel).
At the location of C2, the shock and the magnetopause moved back across the spacecraft and it re-entered the magnetosphere for several seconds at 18:15:20~UT. The other spacecraft stayed in the supersonic jet for over a minute moving gradually back into normal sheath-type plasma.
This transition can be seen in the ion velocity distributions (not shown): the narrow ($\sim$1 MK) distribution of the jet was slowly replaced by a warmer, symmetric quasi-Maxwellian after 18:16~UT.

While in the jet, Cluster observed a gradual increase in both plasma density and magnetic field: from low values of 7~cm$^{-3}$ and 8~nT at the beginning of the jet to very high values of 22~cm$^{-3}$ and 30~nT at the end (see Figure~\ref{fig:V}, second panel for C1 ion density
). Consequently, the dynamic pressure in the jet increased to over 6~nPa, as compared to the nominal pressure of 1~nPa. 
These enhancements were accompanied by a substantial deflection of the bulk flow from its nominal direction, as illustrated by the third panel of Figure~\ref{fig:V}.

\begin{figure}
\includegraphics[width = 0.3\textwidth]{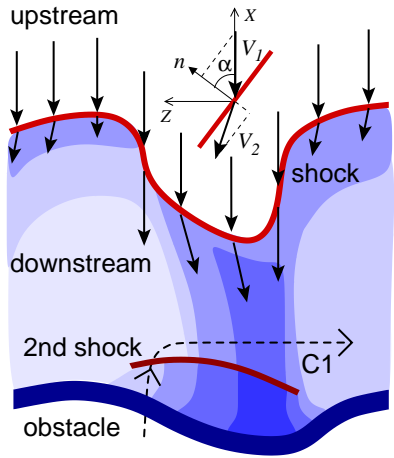}
\caption{Illustration of the effect of local shock curvature. The variation of the plasma number density in the downstream region is illustrated by the shading: dark blue indicates density enhancement, light blue indicates density depletion. The trajectory of C1 in the reference frame moving with the ripple is sketched with the dashed line. The inset details the flow deflection when $\textbf{V}_1$ is not parallel to $\textbf{n}$.\label{fig:cartoon}}
\end{figure}

\textit{Interpretation.---}
We propose the following mechanism to explain the formation of the jet:
First, consider an oblique shock with radial upstream conditions ($\textbf{V}_{1} \parallel \textbf{B}_{1}$) as illustrated in the inset of Figure~\ref{fig:cartoon}. The Rankine-Hugoniot jump conditions for high ${M_\mathrm{A}}$ give $V_{1n} = rV_{2n}$ and $V_{1t} \sim V_{2t}$, where $r$ is the shock compression ratio.

We then consider the streamlines of plasma flow across a curved high  ${M_\mathrm{A}}$ shock as illustrated in Figure~\ref{fig:cartoon}.
We infer, based on the analysis of the 
observations as will be discussed below, that the scale of the 
shock ripple under consideration is of the order of the spacecraft separation: $50-100$ ion inertial lengths, $7000-15000~\mathrm{km}, \mathrm{1-3}~R_\mathrm{E}$.
As the shock primarily decelerates the component of the upstream velocity $\textbf{V}_1$ parallel to the shock normal $\textbf{n}$, the shock crossing leads to efficient compression and deceleration in regions where the angle $\alpha$ between $\textbf{V}_1$ and $\textbf{n}$ is small.
Wherever $\alpha$ is large, however, the shock mainly deflects the flow while the plasma speed stays close to the upstream value.
The plasma is still compressed so that the higher density together with the high speed leads to a jet of very high dynamic pressure. 
Furthermore, if the speed $V_2$ of this jet on the downstream side is still super(magneto)sonic in the reference frame of the obstacle, a second shock front forms closer to the obstacle.
In addition, depending on the ripple geometry, the flow behind the shock can converge causing local density enhancements, or diverge causing density depletions.

Let us compare Figure~\ref{fig:cartoon} with the C1 measurements presented in Figure~\ref{fig:V} where, in the third panel, the bulk flow direction is displayed in the ($-Z_\mathrm{GSE}$, $X_\mathrm{GSE}$) plane.
The observed pattern of the supermagnetosonic flow after the secondary shock suggests that there is a ripple in the bow shock similar to the one of the illustration moving in the $\sim Z_\mathrm{GSE}$ direction. This interpretation is supported by the observed density and flow speed profiles. The fourth panel of Figure~\ref{fig:V} shows the upstream angle $\alpha$ for the supermagnetosonic jet calculated from the observations (considering both the downstream and upstream data and taking $r=4$).
During the main velocity deflection, $\alpha\sim-65^{\circ}$.
The flow pattern in the $Y_\mathrm{GSE}$ (not shown) reveals more of the three-dimensional structure of the ripple and will be considered elsewhere.
The observations of C3 are similar, though not identical to C1. Given this and the fact that C2 was outside of the jet, we infer that the lower limit for the scale of the bow shock perturbation is of the order of the spacecraft separation ($\sim$~8000~km, 1.2~$R_\mathrm{E}$).

The ripples we propose to be the source of the high speed jets stem from the unstable nature of collisionless quasi-parallel shocks: reflected ions can stream against the upstream flow and interact with the incident plasma over long distances before returning (if at all) to the shock.
This interaction triggers instabilities and creates waves that steepen into large structures convecting back to the shock front (see \cite{Burgess05}, and the references therein).
The effect is most pronounced when $\textbf{B}_{1}$ and $\textbf{V}_{1}$ are aligned in the coordinate system of the obstacle.

Both observations and simulations have shown that ripples are inherent to quasi-parallel shocks:
Observations of the ion reflection on the upstream side of the Earth's bow shock \cite{Onsager90} indicated that the direction of $\textbf{n}$ varies when the shock is quasi-parallel at the subsolar point. Such studies on the ion distributions have also shown that, at times, the solar wind does indeed
pass through the shock layer without significant heating \cite{Gosling89}. However, no connection between these two findings was made.
Furthermore, recent multi-spacecraft observations have characterised in detail the 
short, large amplitude magnetic structures (SLAMS) \cite{Lucek02,Lucek08} convecting in the upstream of Earth's quasi-parallel bow shock towards the shock front. SLAMS have a scale size up to 1~$R_\mathrm{E}$ comparable to the ripples discussed here.
In addition, measurements showed signatures that the shock transition itself is narrow consisting of only one to a few SLAMS.
The roughness of the parallel part of
the shock front due to SLAMS is clearly seen in the bow shock simulations of, e.g., Blanco-Cano \textit{et al.} \cite{Blanco-Cano09}.

\textit{Discussion and Conclusions.---}
In previous studies Savin \textit{et al.} \cite{Savin08} have found several jets with high kinetic energy density ($\frac{1}{2}\rho V^2$), 
of which 33 jets had an energy density larger than 10~keV/cm$^{3}$ (compared with 19~keV/cm$^{3}$ in this event).
Nemecek \textit{et al.} \cite{Nemecek98} have also reported what they call transient ion flux enhancements, with fluxes of $6\times10^{8}$/cm$^{2}$s ($\sim8\times10^{8}$/cm$^{2}$s in this event), during intervals of radial interplanetary magnetic field.
Both transient flux enhancements and high kinetic energy jets have properties similar to the jets reported here.
Neither Nemecek \textit{et al.} nor Savin \textit{et al.} could identify a clear source for the jets, but they could rule out, e.g., reconnection.
Here, we propose a generation mechanism for high speed jets in the sheath that is in agreement with the measurements presented in this Letter and those reported in previous studies.
Naturally, we cannot ascertain that all of the previously reported jets stem from the same shock geometry related origin.

It has become evident that shocks are more structured than was previously recognized, so that  a conventional plane wave description is not sufficient. In fact, the mechanism proposed in this Letter for spatial structuring of the downstream is valid for all rippled shocks regardless of magnetic field obliquity, provided that the Mach number is high.  As Voyager 1 and 2 crossed the heliospheric termination shock \cite{Jokipii08Nature}, their observations revealed a rippled, supercritical (${M_\mathrm{MS}}$~$\sim$~10) quasi-perpendicular shock \cite{Burlaga08Nature}. Likewise, interplanetary shocks seem to be nonplanar \cite{Neugebauer05} and also oblique ones may be rippling \cite{Krauss-Varban08}. Therefore we expect that the effects of the ripples, including supersonic jets, can be observed behind collisionless shocks in many plasma environments, and especially behind extended, varying shock fronts having quasi-parallel regions.

In astrophysical context, the high speed jets and nonthermal structure can act as seeds for magnetic field amplification and particle acceleration \cite{Giacalone07}, even for smooth upstream plasma.
In magnetospheric context, the jets with their high dynamic pressure provide a previously unidentified source for magnetopause waves during steady solar wind conditions. A locally perturbed magnetopause is consistent with the Cluster measurements of C2 being within the magnetosphere while the other spacecraft were in the jet. In turn, the large magnetopause perturbation can affect the coupled magnetosphere-ionosphere dynamics \cite{Sibeck89}.
Note also that this Letter presents observations of a weak shock within the magnetosheath during steady upstream conditions. Previous studies of discontinuities within the sheath have been related to bow shock interaction with interplanetary shocks (see \cite{Prech08}, and the references therein).


In summary, we propose a generation mechanism for high speed jets in the downstream side of a quasi-parallel shock based on a 
set of multi-point measurements. Quasi-parallel shocks are known to be rippled even during steady upstream conditions. The local curvature changes of the quasi-parallel shock can create fast bulk flows: in the regions where the upstream velocity is quasi-perpendicular to the local shock normal, the shock mainly deflects plasma flow while the speed stays close to the upstream value.
Together with the compression of the plasma, these localized streams can lead to jets with a kinetic energy density that is several times higher than the kinetic energy density in the upstream.

\begin{acknowledgments}
We thank N. Ness and D. J. MacComas for ACE data, R. Lepping and R. Lin for Wind data, and T. Mukai and T. Nagai for Geotail data.
We also thank CDAWeb at NSSDC and the Cluster Active Archive for data access. H. H. thanks M. Andr\'e and E. K. J. Kilpua for useful comments.
The work of H. H. is supported by the Vaisala foundation and the M. Ehrnrooth foundation. The work of T. L., K. A., R. V., and M. P. is supported by the Academy of Finland. M. P. is also supported by EU-FP7 ERC starting grant 200141-QuESpace.
\end{acknowledgments}


\end{document}